\documentclass[rsi,graphicx, preprint]{revtex4-1}

\usepackage{graphicx}
\usepackage{subcaption}
\usepackage{caption}
\usepackage{float}
\usepackage{threeparttable}
\usepackage[font=small]{caption}
\usepackage{booktabs}
\usepackage{tabularx}
\usepackage{makecell}

\draft 

\begin{document}


\title{High-voltage generation system for a traveling-wave Stark decelerator} 



\author{Lucas van Sloten}
\author{Leo Huisman}
\author{Steven Hoekstra}
\affiliation{Van Swinderen Institute for Particle Physics and Gravity, University of Groningen, The Netherlands}
\affiliation{Nikhef, National Institute for Subatomic Physics, Amsterdam, The Netherlands}


\date{\today}

\begin{abstract}
In this paper we describe the high-voltage generation system we have developed for a traveling-wave Stark decelerator (TWSD). The TWSD can reduce the forward velocity of a molecular beam of heavy neutral polar molecules such as strontium monofluoride (SrF) and barium monofluoride (BaF) from $\sim$ 200 m/s down to $\sim$ 6 m/s. The main motivation for the development of this device is the increased sensitivity from precision spectroscopy of the decelerated molecules to test fundamental physics. The high-voltage generation system can produce eight pulsed sinusoidal waveforms with a maximum amplitude of 10 kV and a linear frequency sweep from 16.7 kHz down to 500 Hz over the span of 40 ms at a repetition rate of 10 Hz. The eight waveforms are phase-offset to each other by 45 degrees. To slow down the heavy molecules, the decelerator is required to have a length of $\sim$ 4 m, which results in a significant capacitive coupling between adjacent channels of $\sim$ 160 pF. As a consequence, the control and stability of the waveforms is extra challenging. We designed a method that compensates for the frequency-dependent coupling between the eight channels. Allowing for amplitude and phase-offsets that do not deviate more than 1\% and 2 degrees, respectively, from their design values during the frequency sweep. The system outperforms commercially available options in terms of stability, output voltage amplitude, cost and ease of maintenance. This approach is also relevant for other fields where precise control of high-voltage waveforms is required, such as particle accelerator physics, plasma physics and mass spectroscopy \cite{Lewandowski.2024}\cite{Villegas.2024}\cite{Zhao.2024}.

\end{abstract}


\maketitle

\section{Introduction}

\subsection{Stark decelerators}
The production of a slow beam of molecules is of interest for high-precision experiments that probe fundamental physics~\cite{collaboration.2018}~\cite{DeMille.2024}. Stark decelerators utilize the Stark shifts of molecules resulting from the electric fields within the device~\cite{Meerakker.2012}. The process works for molecules in a so-called low-field seeking state, which characterizes molecules that have an energy minimum in regions of low electric field strength. By generating local electric field minima, the molecules can be controlled by means of the electric field. To generate the required electric fields, high-voltage is applied to neighboring electrodes. Two main decelerator geometries can be distinguished. The conventional approach uses a crossed-pin electrode geometry called pulsed-pin Stark deceleration (PPSD). In this approach high-voltage switches are used to apply two voltage configurations to the pin electrodes. While the PPSD is relatively easy to implement using commercially available electronics, it suffers from well-documented molecular losses \cite{Sebastiaan.2006}~\cite{Sawyer.2008}. The second approach called traveling-wave Stark deceleration (TWSD) uses a series of ring-shaped electrodes through which the molecular beam travels \cite{Osterwalder.2010}\cite{Sreekanth.2016}\cite{Meek.Osterwalder.2011}\cite{Bulleid.2012}. By applying high-voltage sinusoidal waveforms to the ring electrodes, potential traps are formed. These traps move longitudinally through the decelerator at a speed determined by the frequency of the applied waveforms. By progressively lowering the frequency of the high-voltage waveforms as the molecules travel through the decelerator, the longitudinal beam velocity can be greatly reduced, or the molecules can even be brought to a complete stop. While the TWSD is inherently more stable than the PPSD, it is nonetheless subject to molecular losses. One of the loss mechanisms for a TWSD arises from non-adiabatic transitions to untrappable quantum states near the trap center due the degeneracy of these states at low electric field strength~\cite{Meek.2011}. Furthermore, the requirements on the high-voltage source are significantly greater. Precise control of the waveform parameters is crucial for the effective deceleration and the suppression of molecular losses. In this work the development of such a high-voltage source is described. A particular feature of our application is the required length of the decelerator. A decelerator length of 3 to 4.5 meters (depending on the initial velocity, voltage amplitude and the molecular state~\cite{Berg.Hoekstra.2012}) is required to slow down the relatively heavy BaF molecules. Compared to shorter decelerators \cite{Meek.Osterwalder.2011}, this results in a significant capacitive coupling between adjacent channels and a large overall capacitive load. As a consequence, the control and stability of the waveforms is extra challenging. 

The present paper discusses the development of a high-voltage source to operate a TWSD. 
The paper has the following structure: Section (\ref{sec:mech_design}) provides a general overview of the decelerator design, followed by section (\ref{sec:options_generation}), which outlines the various means of high-voltage waveform generation. Section (\ref{sec:requirements}) details the requirements for the voltage source, and (\ref{sec:overview}) presents an overview of the experimental setup. 
Section (\ref{sec:methods}) describes the implementation in detail. This includes the generation of the waveforms (\ref{sec:waveform_generation}), the audio amplifiers (\ref{sec:amplification_filters}), the development of custom-made transformers (\ref{sec:transformers}), and the feedback system (\ref{sec:feedback_system}). The main results are presented in (\ref{sec:results}), starting with transformer performance (\ref{sec:Bodeplot_all}), followed by waveform optimization at 1 kV and scaling up to 5 kV (\ref{sec:1kv}). Section (\ref{sec:waveform_fidelity}) evaluates the waveform fidelity, and section (\ref{sec:10kv}) presents the results for operation at 10 kV. A discussion of the results and system limitations is provided in section (\ref{sec:discussion}). Finally, the main conclusions and future outlook are summarized in section (\ref{sec:conclusions_outlook}).

\subsection{Mechanical design of the decelerator}
\label{sec:mech_design}

\begin{figure}[H]
    \centering
    \begin{subfigure}[b]{0.4\textwidth}
        \centering
        \includegraphics[width=\textwidth]{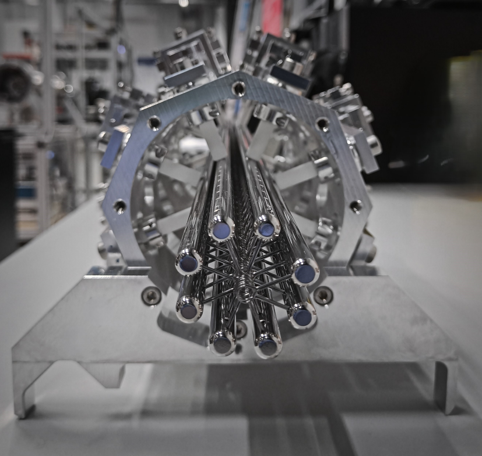} 
        \caption{}
        \label{fig:subfig-a}
    \end{subfigure}
    \hfill
    \begin{subfigure}[b]{0.55\textwidth}
        \centering
        \includegraphics[width=\textwidth]{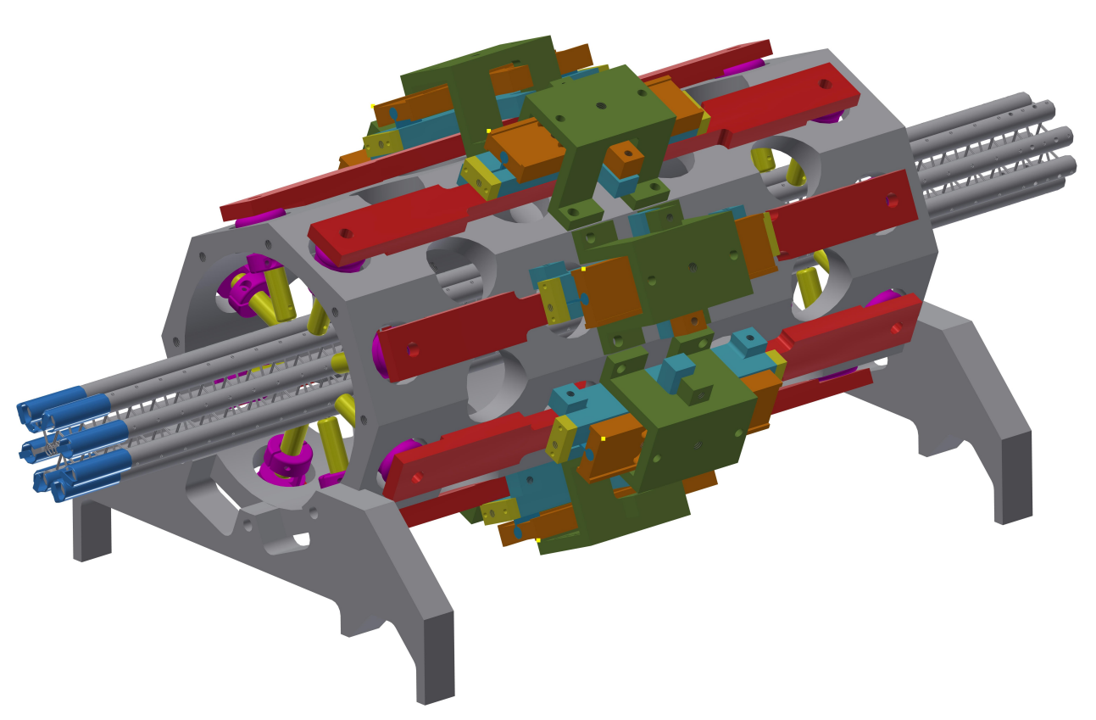} 
        \caption{}
        \label{fig:subfig-b}
    \end{subfigure}
    \caption{(a) Picture of the front view of one of the decelerator modules. (b) Isometric side view of the module showing the mounting structure used to hold and align the electrode rods.}
    \label{fig:modules}
\end{figure}

The design of the decelerator is very similar to the setup used in Meek \textit{et al}. \cite{Meek.Osterwalder.2011}.
The decelerator consists of eight sets of ring-shaped electrodes with an inner radius of 2 mm, made of tantalum wire with a diameter of 0.6 mm. The center-to-center separation between adjacent rings is 1.5 mm. Each set of rings is connected to a 8 mm thick stainless steel rod, which is mounted in an octagonal configuration as shown in Figure~\ref{fig:modules}.
The decelerator has a modular design, with each module measuring 504 mm in length and containing 336 ring electrodes. While the decelerator described in Meek \textit{et al} consists of a single module mounted vertically, the decelerator described in this work consists of six modules that are oriented horizontally,  resulting in a total length of 3024 mm and 2016 ring electrodes. The modules are mounted on top of a rail located inside the vacuum chamber that forms the main body of the decelerator. The rail ensures that all modules are properly aligned in the transverse direction. Each rod and therefore the connected set of rings can be individually adjusted for each module using adjustment screws. The modules are connected to each other by spring-loaded copper sleeves at the ends of each rod.

The high-voltage capabilities are tested for each module separately by applying a DC voltage to one of the rods and measuring the current going to the neighboring rods and to ground. A 50 M$\Omega$ resistor is used in series to limit the current, thereby preventing damage from any discharges that may occur. If the currents are small ($\sim \mu A$) and stable for all rods, then the module is considered acceptable. Otherwise the module can be conditioned by applying a voltage amplitude just below the point at which the currents become unstable or too large. By applying this voltage for several minutes, the performance of the module can be improved depending on the nature of the cause of the performance issues. For a sinusoidal waveform with an amplitude of 10 kV and a phase offset of 45 degrees w.r.t. the neighboring channel, the maximum potential difference between any two adjacent rods is 7.7 kV. To ensure safe operation, each module should be rated to withstand at least 8 kV between adjacent rods before large or unstable currents occur. 

\newpage

\begin{figure}[H]
    \centering
    \includegraphics[width=0.5\linewidth]{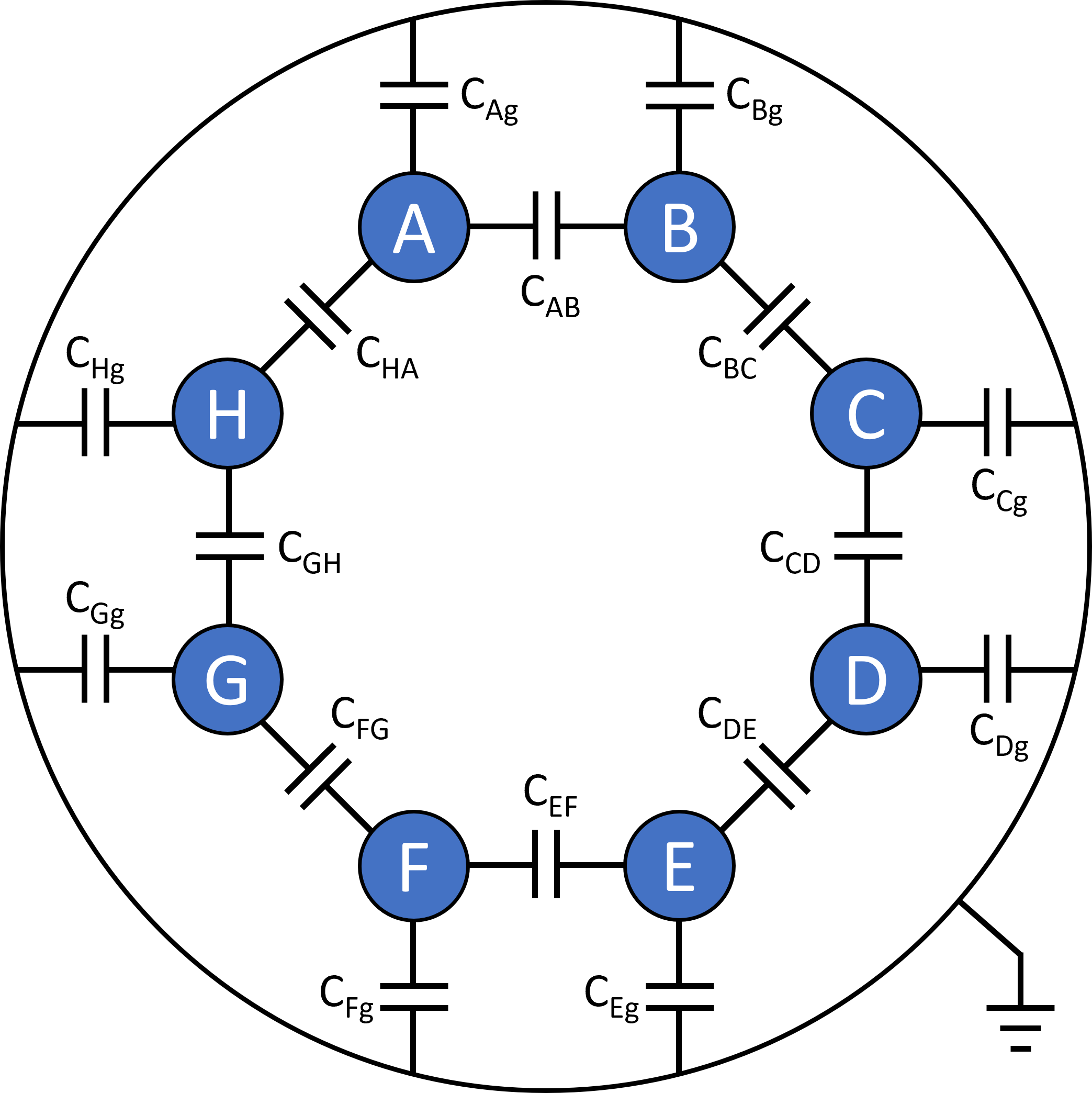}
    \caption{Equivalent circuit of the decelerator. The rods are labeled A through H. For a 4.5 m long decelerator, the capacitance between each pair of adjacent channels is determined to be $\sim$ 160 pF and between each channel and ground $\sim$ 57 pF \cite{Zapara.2019}.}
    \label{decelerator_equivalent_circuit}
    
\end{figure}

The large number of rings that are in close proximity to each other result in a significant capacitance between adjacent channels. Additionally, the mounting structure and vacuum chamber introduce capacitance between each channel and ground. Figure \ref{decelerator_equivalent_circuit} shows the equivalent circuit of the decelerator.

The effective load experienced by the voltage source is caused by the capacitance of the decelerator and is estimated to be $\sim$ 151 pF for a 4.5 meter long decelerator.

Due to the capacitive coupling between adjacent channels, any change in voltage applied to one channel influences the voltage on all other channels. This makes precise control over the eight waveforms challenging. To overcome these difficulties, a feedback system has been developed, which is described in section (\ref{sec:feedback_system}).
The modular design of the decelerator allows its length to be adjusted. While the high-voltage source is designed to operate with the maximum decelerator length of 4.5 meter, measurements described in this work have been performed on a 3 meter configuration consisting of six modules. This reduces the capacitance by a factor of $2/3$, resulting in a capacitance between adjacent channels of 107 pF and an effective capacitive load of 101 pF. However single-channel tests have also been performed on a 200 pF dummy load as described in section (\ref{sec:10kv}).

\subsection{Options for high-voltage waveform generation}
\label{sec:options_generation}
The generation of the high-voltage waveforms required for the operation of a TWSD is usually accomplished by either high-voltage amplifiers, as in the work of Fitch \textit{et al.} \cite{Shyur.Lewandowski.2018ggo}, or by using transformers. High-voltage amplifiers are electronically more complex, but generally have a larger bandwidth and can be used to bring the molecules to a complete standstill by applying DC voltages to the electrodes. In contrast, transformers usually have a limited bandwidth and can not be used to bring the molecules to a complete standstill. However, transformers offer a relatively inexpensive alternative and have been successfully implemented to slow down molecules, as demonstrated by Meek \textit{et al.}~\cite{Meek.Osterwalder.2011}, where CO molecules were decelerated from 300 m/s to 180 m/s.\\
The decelerator described in the present paper has previously been used to slow down SrF molecules in the N=1 rotational state from 190 m/s to a complete standstill as described in Aggarwal \textit{et al.} \cite{Aggarwal.2021} and BaF in the N=1 state from 200 m/s to 150 m/s in \cite{Touwen.2024}, using custom-made Trek high-voltage amplifiers (model PD10039). The performance of these amplifiers is therefore used as a benchmark for evaluating the performance of the present transformer-based system. The main specifications of the Trek amplifiers and the systems used by Meek \textit{et al.} and Fitch \textit{et al.} can be found in table 1. The voltage required to achieve maximum trap depth depends on the specific molecule, its quantum state and the geometry of the decelerator. For BaF in the N=1 state this is approximately 5 kV. Higher voltages would cause the molecules in this state to transition to high-field seeking states, resulting in a reduced capture volume. In the N=2 state however this maximum is approximately 10 kV, allowing for the formation of deeper traps. This significantly increases the phase space acceptance of the decelerator resulting in a factor $\sim4$ increase in the number of molecules \cite{Touwen.2024}. The switch to a transformer-based system is made to be able to reach this higher voltage and to reduce the dependence on external suppliers, since it allows key components to be built in-house. 

\begin{table}[tbp]
\centering

\begin{threeparttable}
\caption{Main specifications of the Trek amplifiers, the present transformer-based system, the system used by Meek \textit{et al.}, and the system used by Fitch \textit{et al.}.}

\begin{tabular}{|c|c|c|c|c|}
\hline
 & Meek \textit{et al.} \cite{Meek.Osterwalder.2011} 
 & Fitch \textit{et al.} \cite{Shyur.Lewandowski.2018ggo} 
 & Trek amplifier \cite{Zapara.2019} 
 & Present system \\
\hline
Type
 & \makecell{Transformer\\[-3.5mm]based}
 & \makecell{Solid-state\\[-3.5mm]amplifier}
 & \makecell{Solid-state\\[-3.5mm]amplifier}
 & \makecell{Transformer\\[-3.5mm]based} \\ 
\hline
Max output voltage
 & $\pm$ 10 $\mathrm{kV}$
 & $\pm$ 10 $\mathrm{kV}$
 & $\pm$ 5 $\mathrm{kV}$
 & $\pm$ 10 $\mathrm{kV}$\\
\hline
Max output current
 & unspecified
 & $\pm$ 1.5 $\mathrm{A}$
 & $\pm$ 500 $\mathrm{mA}$
 & $\pm$ 250 $\mathrm{mA}$ \\
\hline
Minimum frequency
 & 12 kHz
 & 0 Hz
 & 0 Hz
 & 500 Hz\\
\hline
Maximum frequency
 & 30 kHz
 & 30 kHz
 & 30 kHz
 & 20 kHz\\
\hline
Voltage gain
 & 250 $V/V$
 & 12000 $V/V$
 & 1000 $V/V$
 & 100 $V/V$ \\
\hline
Total harmonic distortion
 & unspecified
 & $< 1.4\%$\tnote{a}
 & $< 2\%$\tnote{b}
 & $< 0.25\%$\tnote{c}  \\
\hline
\end{tabular}

\begin{tablenotes}
 \footnotesize
  \setlength{\itemsep}{0pt}
  \setlength{\parsep}{0pt}
  \setlength{\parskip}{-5pt}
  \item[a] Measured at 10 kHz with a 500 pF load.
  \item[b] Determined at 28.75 kHz.
  \item[c] Measured at 10 kHz at 5 kV on the 3m-long decelerator.
\end{tablenotes}
\end{threeparttable}

\end{table}

\newpage

\section{System requirements}
\label{sec:requirements}
\begin{figure}[H]
    \centering
    \includegraphics[width=1\linewidth]{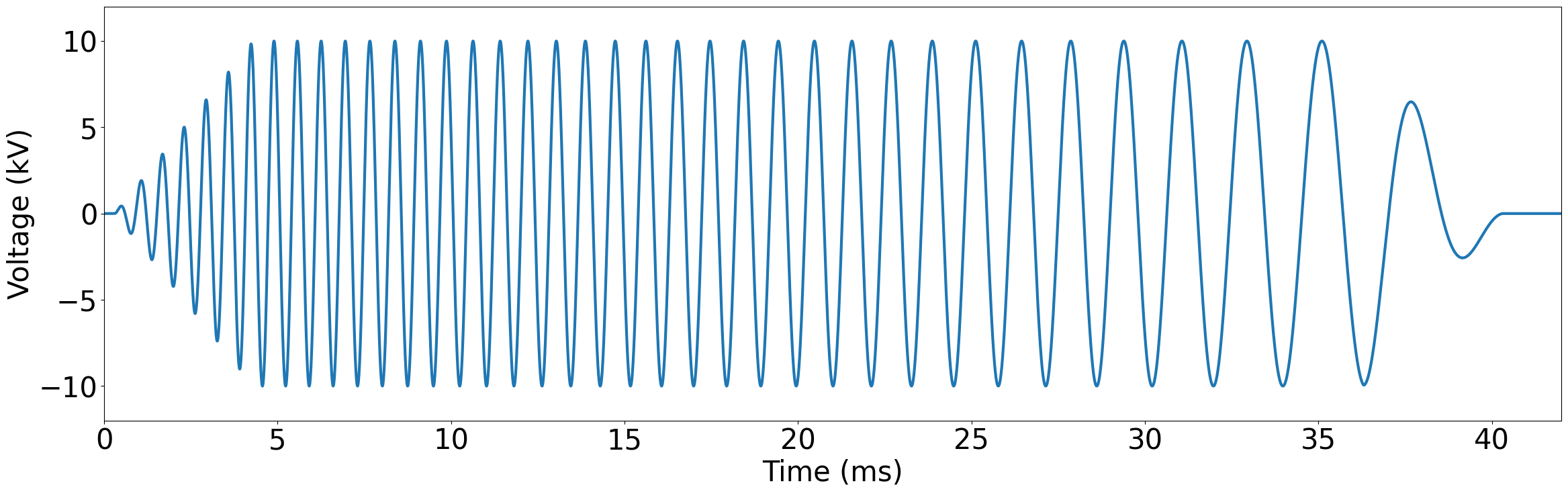}
    \caption{Illustration of a single optimal output waveform. For clarity, the frequency is reduced by a factor of 10 and the ramp-up and ramp-down duration has been set to 10\%.}
    \label{fig:waveform_model}
\end{figure}

The goal is to generate eight channels of pulsed sinusoidal waveforms with a voltage amplitude up to 10 kV on a capacitive load of 151 pF, with a phase offset of 45 degrees between adjacent channels. The waveform frequency is a linear sweep from 16.7 kHz to 2.5 kHz during 40 ms, at a repetition rate of 10 Hz.  The initial frequency is determined by the longitudinal velocity of the molecules as they exit the two-stage cryogenic buffer gas source and corresponds to about 200 m/s. The frequency at the end of the sweep is determined by the desired end velocity and corresponds to about 30 m/s. This final velocity is ideal for the interaction zone used in the NL-eEDM experiment \cite{collaboration.2018}. By sweeping down to even lower frequencies of $\sim$ 800 Hz, final beam velocities can be achieved that allow for direct loading into electric or optical traps \cite{Schellenberg.2026}.  The pulse duration is determined by the average velocity and the length of the decelerator. The waveforms have a linear ramp-up and ramp-down period at the beginning and end of the pulse, respectively, to prevent current spikes due to capacitive effects. These ramp-up and ramp-down periods are usually chosen to last for 1\% of the total pulse length. Figure~\ref{fig:waveform_model} shows an illustration of a single ideal waveform.

Precise control of these eight high-voltage waveforms is crucial for effective deceleration and minimization of molecular losses. Less-than-ideal waveforms and ring misalignment cause distortions of the trap potentials. A quantitative analyses of the molecular losses caused by these distortions is challenging although previous work in this regard does provide some insight \cite{Zapara.2019}\cite{Touwen.2024}\cite{Berg.2015}\cite{Sreekanth.2023}. From this it can be concluded that a total harmonic distortion (THD) of less than $2\%$ in the relevant range of voltage amplitudes and frequencies is sufficient. To further investigate the effects of imperfections in the waveforms, simulations will be done using measured high-voltage waveforms as input. Although the exact required tolerances are unknown as of this writing, an amplitude deviating of less than 1\% and a phase difference within 2 degrees of the desired value are not expected to result in a meaningful loss of molecules and these values are therefore set as the goal. 

\newpage

\section{Design overview}
\label{sec:overview}
\begin{figure}[H]
    \centering
    \includegraphics[width=0.9\linewidth]{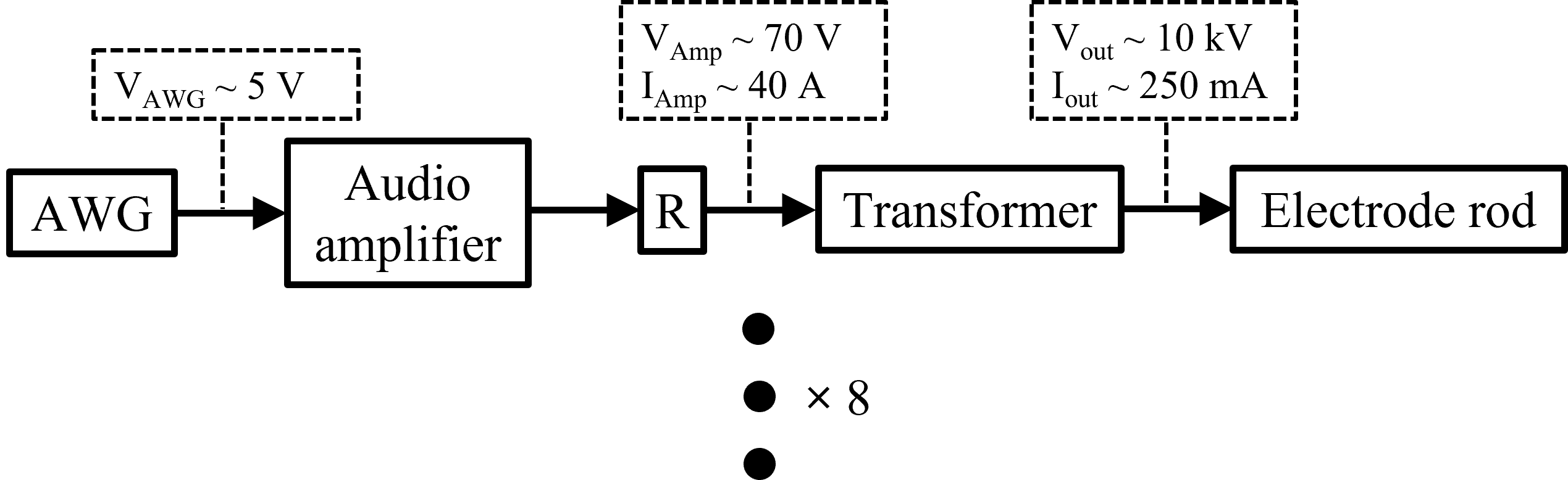}
    \caption{Schematic overview of the setup. The sequence is repeated for each channel. The dashed boxes show approximate values for the voltage and current amplitudes in that location of the setup when using a test load of 200 pF at 10 kV at the starting frequency of 16.7 kHz. For clarity, the capacitive coupling between adjacent channels and between each channel and ground is not shown.}
    \label{fig:schematic_overview}
    
\end{figure}

A key aim of the electronics is to enable the deceleration of BaF in the N=2 rotational state, while also considering compatibility with other potentially interesting molecular species such as barium monohydroxide (BaOH) \cite{Bause.2025}. Maximum trap depth for BaF in the N=2 rotational state for our decelerator geometry is achieved at a voltage amplitude of 10 kV. To produce this voltage, a transformer-based approach is employed. The eight waveforms are generated using arbitrary waveform generators (AWG). These low-power signals form the inputs of eight audio amplifiers that can each deliver a peak power of up to 3000 W to its connected transformer. A resistor is placed between each audio amplifier and transformer to ensure proper functioning of the amplifier. To compensate for the frequency-dependent response of the system, a feedback system is utilized. By predistorting the waveforms generated by the AWG's, the unwanted distortions caused by the system are compensated for. Figure~\ref{fig:schematic_overview} shows a schematic overview of the experimental setup. More details about the setup are found below.

\newpage

\section{Experimental methods}
\label{sec:methods}
\subsection{Waveform generation}
\label{sec:waveform_generation}

The eight waveforms are generated and measured using two Moku Pro devices \cite{MokuPro}, each equipped with four input and four output channels. In this experiment, these multipurpose devices function as both arbitrary waveform generators and as oscilloscopes. Both devices are externally triggered by the same 10 Hz trigger pulse. Each waveform can be measured and modulated separately, which allows for predistortion of the generated waveforms to compensate for unwanted distortions caused by the system.         

\subsection{Voltage amplification and digital filters}
\label{sec:amplification_filters}
To drive the capacitive load of the decelerator and reach the desired amplitude, a considerable amount of power is required. For this purpose, Behringer NX3000D audio amplifiers \cite{NX3000D} are utilized, which have a manufacturer-rated total peak power of 3000 W. The amplifiers are used in bridge mode which combines the two channels of each amplifier in order to provide sufficient current, such that the required amplitude of 10 kV can be reached. The amplifiers feature digital filters that can be used to maintain a constant amplitude of the signal during the sweep, although we chose to develop the more versatile feedback system described below. The amplifiers delay the signal by on average $600\, \mu s$; the differences in these delays between the amplifiers is compensated for by the feedback system.

\subsection{High-voltage transformers}
\label{sec:transformers}
We have designed and manufactured step-up transformers to produce the high voltage required for the operation of the decelerator. These transformers must be able to reliably produce amplitudes of up to 10 kV across the operational frequency range of 2-20 kHz and maintain this amplitude throughout the decelerator operation. At an operating voltage of 10 kV and a frequency of 20 kHz this corresponds to a required current of approximately 190 mA that the transformer needs to deliver to the effective load of 151 pF of the 4.5 meter long decelerator. The transformers are designed such that their two resonance frequencies are outside this operating bandwidth, to minimize amplitude and phase fluctuations across the frequency range. These resonance frequencies arise from the interaction between the leakage inductance and the parasitic capacitance of the transformer. Specifically, the components of the equivalent circuit of a transformer form both a parallel resonance and a series resonance frequency. The physical origin and impact of these two resonances differ. The parallel resonance is primarily associated with the magnetization inductance of the core in parallel with the combined parasitic and load capacitance. At this resonance, the inductive and capacitive currents cancel, resulting in a sharp increase in the input impedance and a rapid phase change between input voltage and input current, while the voltage transfer to the secondary coils is only weakly affected. In contrast, the series resonance arises from the leakage inductance in series with the parasitic and load capacitance. At this resonance the series impedance is minimized, leading to enhanced voltage transfer and a phase change between the input and output voltages. 

\begin{figure}[H]
    \centering
    \includegraphics[width=0.75\linewidth]{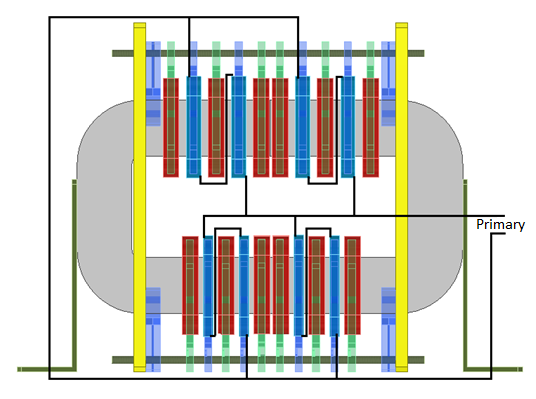}
    \caption{Schematic side view of a transformer. The primary and secondary coils are denoted in blue and red, respectively. Four sets of two primary coils each are connected in parallel, as indicated by the black lines, while the secondary coils are all connected in series (the connections for the secondary coils are not shown). The core material is indicated with grey, and the yellow parts are the mounting structure. }
    \label{fig:schematic_transformer}
    
\end{figure}

Based on these requirements, we have arrived at the design shown in Figure~\ref{fig:schematic_transformer}. The primary side of the transformer consists of 8 coils, divided in 4 pairs of two in-series-connected coils each. The 4 pairs are connected in parallel to spread the large input current over multiple coils. Since the voltage across each pair is the same, the effective number of turns on the primary side is equal to the number of turns in a single pair. The secondary side consists of 12 coils which are all connected in series. The primary and secondary coils are alternatingly mounted on a metal core. The coils and their supporting structure are made in-house to reduce the dependency on external suppliers and thereby the cost and to meet the aforementioned requirements. Specifically, the parasitic inductance and capacitance of the transformer need to be low enough to ensure that the resonance frequencies are outside of the operational frequency range. Previous versions of the transformer were made in a more conventional way by winding the secondary coil over the primary coil. This caused the series resonance frequency to fall within the operational frequency range due to excessive leakage inductance from the secondary coil. Based on a design suggestion by ir. O.C. Dermois, the primary and secondary coils were mounted alternatingly on the core, which sufficiently reduced the leakage inductance. The coils are made from 0.3 mm diameter insulated wire (Magnebond CAB-200). For the primary coils 16 of these wires are wound in parallel on each coil and for the secondary coils only a single wire is used. The cores are provided by Acal BFi and are made of laminated nanocrystalline iron (kOr 120) which has a high magnetic permeability ($3 \times 10^4$ to $1.2 \times 10^5$, depending on frequency) to guide and confine the magnetic fields. Within a transformer’s core, several loss mechanisms are at play. The changing magnetic field introduces Eddy currents in the core. Due to the electrical resistivity of the core’s material, these currents produce heat and are therefore a source of energy loss. To minimize these Eddy currents, the core is made of a large number of sheets that are electrically isolated from each other by an insulating coating. These laminations are 20 $\mu m$ thick and are oriented such that they are parallel to the magnetic field. This confines the Eddy currents in narrow loops, thereby greatly reducing them.  The main specifications of the transformers are summarized in the Table 2. 

\begin{table}[H]
\centering
\caption{Main transformer specifications. The voltages, currents and resonance frequencies are determined for a load of 200 pF at a driving frequency of 20 kHz.}
\begin{tabular}{|c|c|c|}
\hline

Max output voltage & $\pm 10\,\mathrm{kV}$ \\
\hline
Max output current & $\pm 250\,\mathrm{mA}$ \\
\hline
Max input voltage & $\pm$70 V \\
\hline
Max input current & $\pm$40 A \\
\hline
Low resonance frequency & 740 Hz \\
\hline
High resonance frequency & 32 kHz \\
\hline
Core material & Nanocrystalline iron (kOr 120) \\
\hline
Total number of primary coils & 8 (4 sets of 2)\\
\hline
Number of turns per primary coil & 15 \\
\hline
Total number of primary turns & 120 \\
\hline
Effective number of primary turns & 30 \\
\hline
Number of secondary coils & 12 \\
\hline
Number of turns per secondary coil & 250 \\
\hline
Total number of secondary turns &3100 \\
\hline
Effective turns ratio & 1:100 \\
\hline
\end{tabular}
\end{table}

With the 30 primary turns a total number of 3100 secondary turns are needed to achieve an effective turn ratio of 1:100, thereby compensating for losses.

Due to frequency limitations of the electronics that drive the transformer, the 32 kHz resonance is not measured directly. Instead, a capacitive load of 669 pF is used, which shifts the resonance frequency into an experimentally accessible frequency range. From this measurement, the resonance frequency for a load of 200 pF is inferred. The 740 Hz resonance lies within the bandwidth of the driving electronics and is therefore measured directly.

\newpage

The 32 kHz series resonance does not limit the operation of the decelerator since it is well above the 20 kHz upper limit of the NX3000D audio amplifier. Furthermore, the initial longitudinal velocity of the molecular beam is low enough as to not require frequencies above 17 kHz. 

The lower parallel resonance of 740 Hz does not produce meaningful changes in voltage gain or a significant phase shift between the input and output voltage. Instead, it results in a phase change between the input voltage and the input current, which is not relevant for Stark deceleration. The transformer is designed such that the parallel resonance frequency is outside the operational bandwidth as a conservative design choice, since doing so only puts minimal constraints on the transformer geometry.
At these lower frequencies, transformer core saturation can become the dominant limitation, as waveform distortions arise once the core saturates. Core saturation occurs at a magnetic flux density of 1.2 T \cite{kOr120}. For a sinusoidal waveform with an amplitude of 10 kV, core saturation occurs at a frequency of 467 Hz, corresponding to a final beam velocity of 5.6 m/s. Measurements using a 200 pF test load at 500 Hz show no significant waveform distortion at an voltage amplitude of 10 kV, demonstrating that the core is indeed not yet saturated at that frequency.

The likelihood of discharges within the transformer increases with voltage amplitude and become significant at levels around 6 kV. They are therefore sealed in a gas-tight metal enclosure filled with sulfur hexafluoride (\(\mathrm{SF}_6\)) gas, which has a much higher dielectric strength compared to air. 
Alternative insulation approaches were also explored such as operation under high-vacuum conditions or encapsulation in a high dielectric strength casting resin. However consideration relating to heat exchange, capacitive effects and ease of maintenance ultimately led to the use of \(\mathrm{SF}_6\) gas as the preferred insulation medium.  

Although the transformers are designed such that their resonance frequencies are outside of the frequency domain used in the decelerator, the amplitude of the output voltage of the transformer is not constant over the operating range, as can be seen in Figure 7. To ensure maximum trap depth throughout the frequency sweep, this variation is compensated for by the feedback system as described below.

\newpage

\subsection{Monitoring and feedback system}
\label{sec:feedback_system}

To monitor the output of the transformers while they are connected to the decelerator, voltage dividers are used. These reduce the voltage by a calibrated factor of $\sim$ 1000, which allows the voltage to be safely measured using the oscilloscope function of the Moku Pro devices. In order to predistort the waveforms appropriately, a python-based feedback system has been developed that runs on a computer connected to the Moku Pro devices. The feedback system takes the measured outputs of the transformers and determines the waveforms that the Moku Pro devices need to generate such that the output of the transformers have the desired amplitude and phase. Due to the capacitive coupling between the channels this is done iteratively to ensure convergence of the generated waveforms. To prevent damage to the decelerator and the associated electronics, the voltage between adjacent channels and the voltage between each channel and ground should not exceed certain predetermined thresholds. The feedback system therefore has several build in safety features. The feedback system consists of multiple distinct parts, referred to as modules, described below.

\subsubsection*{Waveform generator} 
\vspace{-0.5em}

The waveform generator takes two inputs for each channel; a lookup table and an amplitude. The lookup table is a time-series of normalized voltage values representing the waveform and the amplitude scales it accordingly. The lookup table consists of a maximum of 65536 data points, which provides sufficient resolution for the envelope and amplitude feedback. However, this resolution is inadequate for precise phase feedback at the highest frequency. For example, for a 40 ms pulse this results in a roughly 4 degrees phase difference between adjacent data points at an initial frequency of 16.7 kHz. The lookup table is therefore not used for the phase feedback. 

\subsubsection*{Amplitude checker} 
\vspace{-0.5em}

The amplitude that the waveform generator takes as an input is checked to ensure that it falls within acceptable bounds. When this amplitude falls outside of the acceptable bounds, the nearest allowed value is used instead.

\subsubsection*{Oscilloscope}
\vspace{-0.5em}

Once an initial set of waveforms has been generated, the output of the transformers is measured using the oscilloscope functionality of the Moku Pro devices. The oscilloscope takes the signals entering the input channels of the Moku Pro devices and digitizes them for further processing. 

\subsubsection*{Safety test}
\vspace{-0.5em}

After each measurement, a safety test is performed. The data from the oscilloscopes is analyzed to determine the largest amplitude across all waveforms, as well as the largest difference between adjacent waveforms. These values are then compared to predetermined values that are deemed safe. If one or both of the measured values are larger than the corresponding predetermined value, the amplitudes of all waveforms are scaled down proportionally based on the value the furthest above the safety threshold. Prior to generating new waveforms, the amplitudes of all waveforms are ramped down to zero in a predetermined number of steps. The new waveforms are then applied, and the amplitudes are ramped back up to the desired value in the same number of steps. At each step the safety test is performed, preventing potential violation of the safety thresholds upon the application of the new waveforms.

\subsubsection*{Envelope feedback}
\vspace{-0.5em}

When the safety test has been passed, the envelopes of the output waveforms are determined. This is done by fitting a polynomial to the peaks of the waveforms between the end of the ramp-up and the beginning of the ramp-down. The degree of the polynomial can be increased for better accuracy at the cost of computational speed. A fourth order polynomial has been found to strike a good balance between the two and has thus been used unless stated otherwise. The difference between the maximum and the minimum of each polynomial is determined. The ratio between this difference and the maximum amplitude of the waveform is referred to as 'envelope distortion'. The largest of this set of values determines which waveform will be updated. The inverse of the polynomial with the largest deviation is determined and multiplied by the corresponding previous input waveform. In this way any distortion of the envelope due to the system is compensated for without the need for detailed knowledge of the behavior of the system. When the envelope distortion is smaller than some predetermined value, no changes are made and the feedback system goes to the amplitude equalizer described below. In the context of the feedback system 'input waveforms' refers to the waveforms produced by the AWG while 'output waveforms' refers to those measured at the output channels of the transformers.

\subsubsection*{Amplitude equalizer}
\vspace{-0.5em}

To match the output waveform amplitudes to a desired target value, the ratio between the target amplitude and the maximum of each measured amplitude is calculated. The waveform with the largest deviation from the target is scaled by the corresponding ratio. When all amplitudes are within some predetermined margin of the target value, no changes are made and the feedback system continues to the next module. 

\subsubsection*{Phase feedback}
\vspace{-0.5em}

Each waveform needs to be 45 degrees phase shifted relative to its adjacent channels. To achieve this, the phase differences between all neighboring channels are calculated, and the pair with the largest deviation from the target 45 degrees is identified. To correct this deviation, one waveform in the pair is phase shifted by applying an appropriate time delay. To measure the waveforms with enough time resolution, only the first millisecond of the waveform is considered when determining the phase. Due to the frequency sweep, any delay added has the greatest effect on the phase at the start of the waveform. Instead of using the look-up table to shift the waveform, the triggering event is used. The AWG produces a single trigger pulse which itself is triggered by the 10 Hz signal from the external oscilloscope. The trigger produced by the AWG is a linear ramp-up from 0 to 1 V in a time equal to a single period of the waveform. Each waveform can then be shifted by changing the voltage at which a trigger event occurs for that waveform. The trigger level of each channel can be adjusted with millivolt precision, allowing for phase adjustments as small as 0.36 degrees.  When all phase differences are within some predetermined margin of the target value, no changes are made and the feedback system continues.

\begin{figure}[H]
    \centering
    \includegraphics[width=0.8\linewidth]{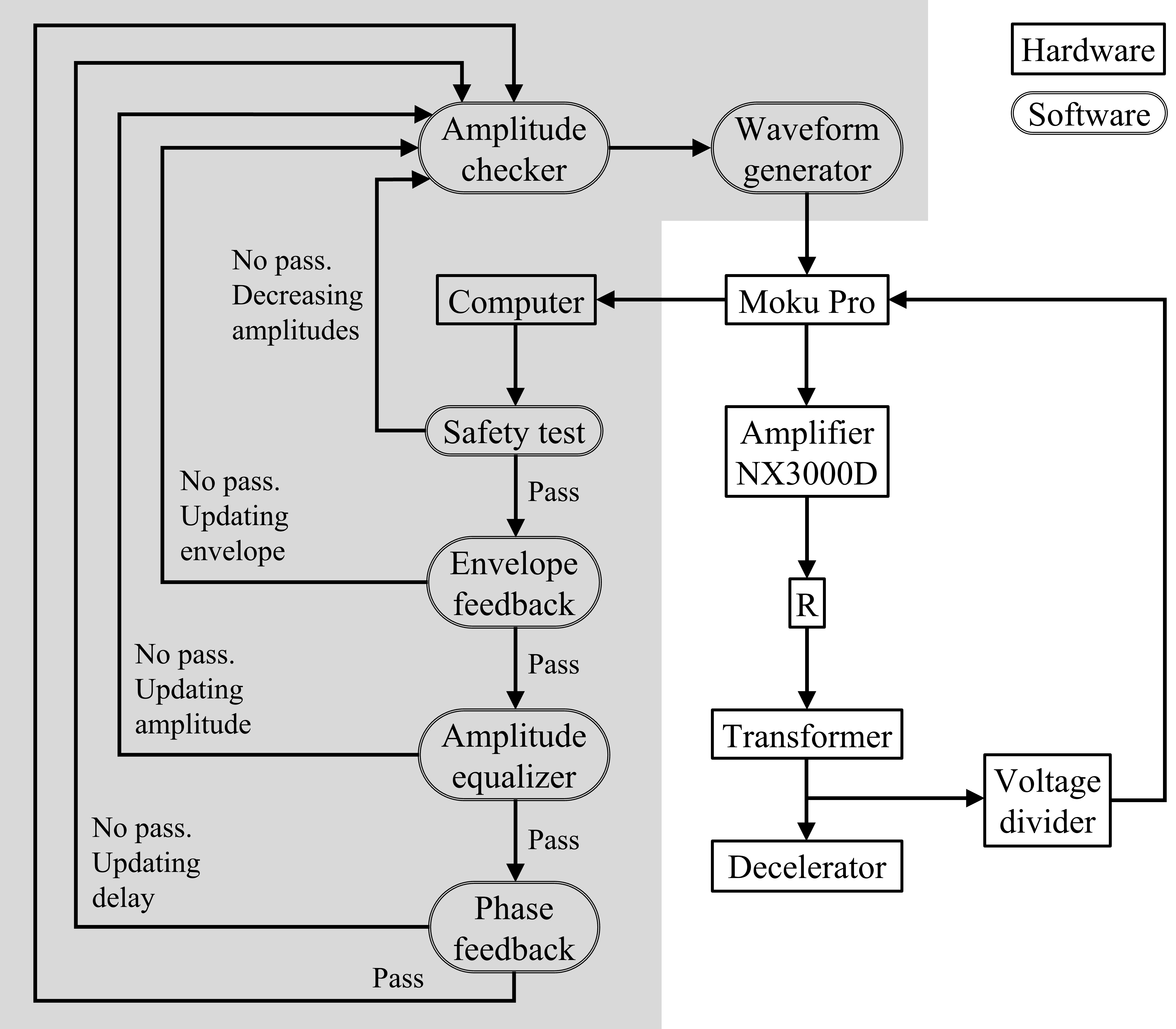}
    \caption{Flowchart of the feedback system represented for a single channel. Everything within the shaded area runs on the computer. In each module, the relevant parameter values are determined. If these are within predetermined acceptable bounds, the system proceeds to the next module indicated with 'Pass'. Otherwise, the module determines the required adjustments to the waveform parameters and the system returns to the amplitude checker indicated with 'No pass'.}
    \label{fig:flowchart_feedback}
    
\end{figure}

Figure~\ref{fig:flowchart_feedback} shows an overview of the feedback system.

To prevent over-optimization at the beginning of the procedure, the acceptable tolerance of the envelope feedback system is linearly scaled down until it reaches the desired final tolerance in a set number of iterations.
When all waveforms are within specs, the feedback system only continues to loop through the safety test. Most of the optimization is done at a relatively low voltage, typically 1 kV. The input waveforms acquired from this process can then be scaled up to get the desired output amplitude, after which only minor adjustments are needed. The complete feedback process takes roughly 30 minutes, although this strongly depends on the settings, particularly the acceptable final tolerances and the number of steps used in the ramp-up and ramp-down sequence when applying new waveforms.

\newpage

\section{Results}
\label{sec:results}

In this section, the main results of the performance of the high-voltage waveform generation system are presented, both on a capacitive test load and the 3-meter-long configuration of the decelerator.

\subsection{Transformer behavior}
\label{sec:Bodeplot_all}

\begin{figure}[H]
    \centering
    \includegraphics[width=1\linewidth]{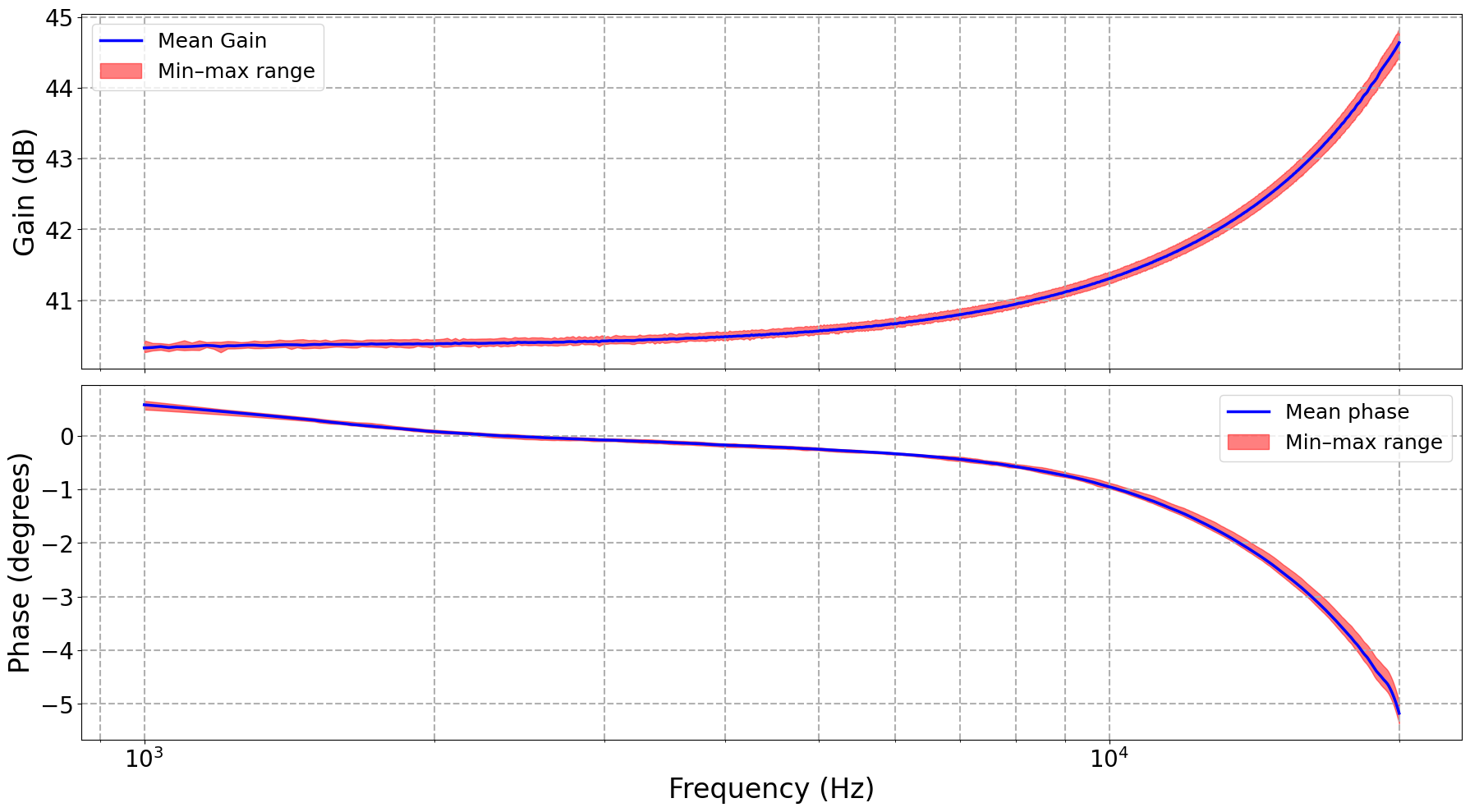}
    \caption{Bode plot averaged over all eight transformers, showing both the voltage gain and phase shift between the input and output of the transformers as a function of frequency. The red bands indicate the minimum and maximum values among the transformers, indicating that all transformers perform nearly identical. A load of 200 pF was used to represent the load of the decelerator.}
    \label{fig:bodeplot}
    
\end{figure}

Figure~\ref{fig:bodeplot} shows the average frequency response of the eight transformers using a 200 pF capacitor to represent the load of the decelerator.

From Figure~\ref{fig:bodeplot} it is clear that a significant phase shift occurs during the sweep. This phase difference is however between the input voltage of the transformer and its output voltage. Since all transformers have near identical phase shifts, the phase difference between the adjacent waveforms caused by the transformers is much smaller.
\newpage

\subsection{Optimization at 1 kV}
\label{sec:1kv}

\begin{figure}[H]
  \centering

  \begin{subfigure}[b]{0.49\textwidth}
    \includegraphics[width=\textwidth]{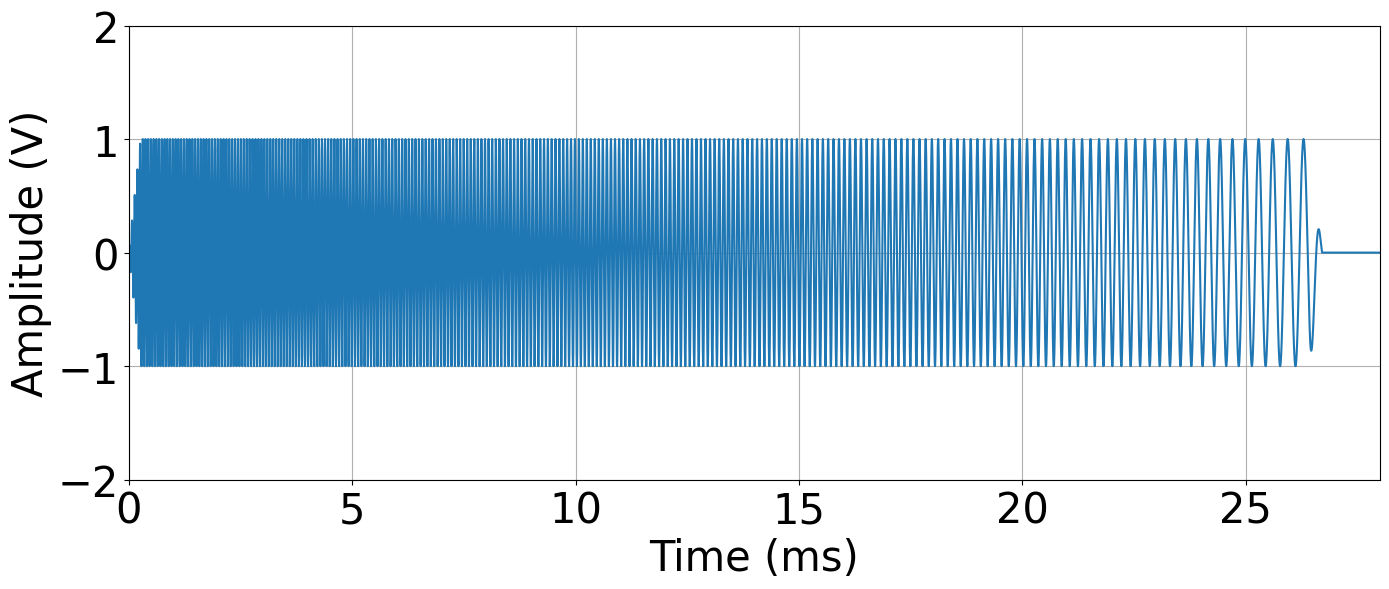}
    \caption{}
    \label{fig:subfig1}
  \end{subfigure}
  \hfill
  \begin{subfigure}[b]{0.49\textwidth}
    \includegraphics[width=\textwidth]{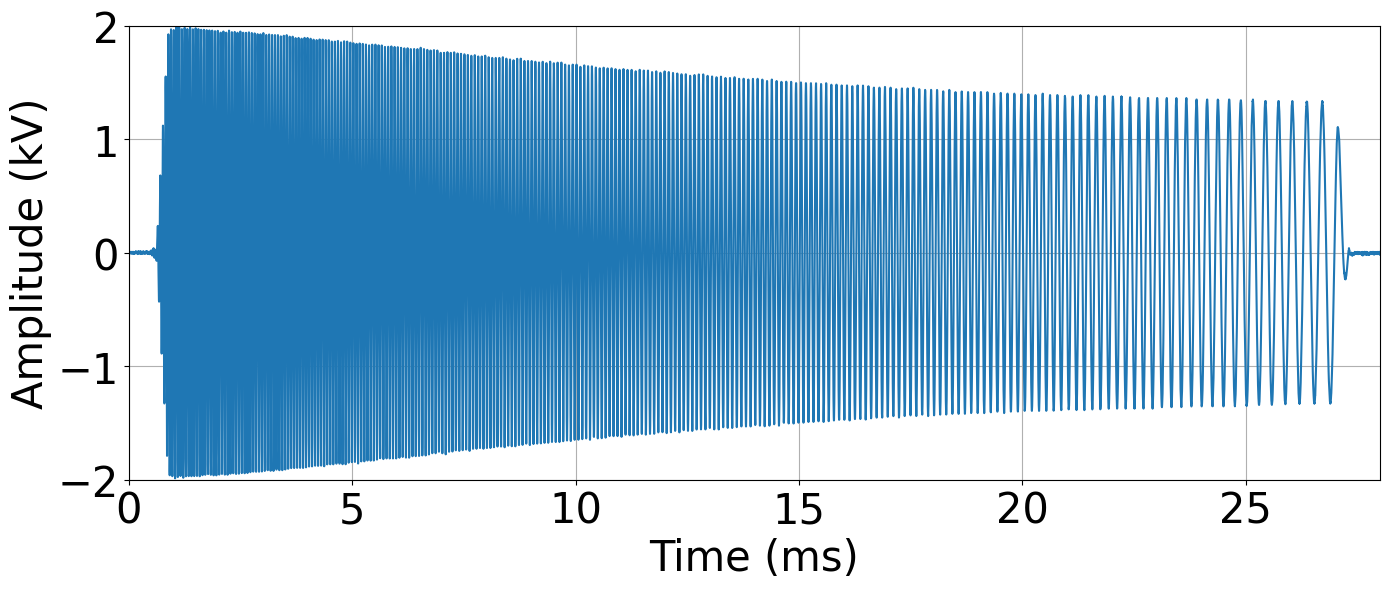}
    \caption{}
    \label{fig:subfig2}
  \end{subfigure}

  \begin{subfigure}[b]{0.49\textwidth}
    \includegraphics[width=\textwidth]{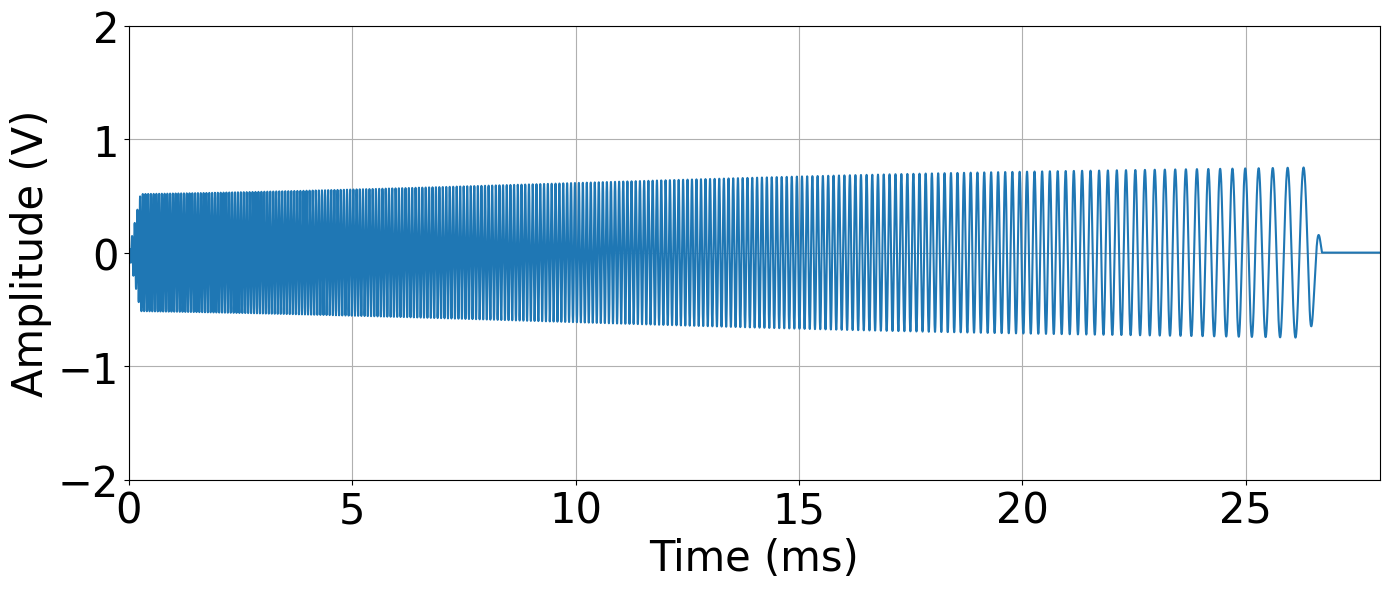}
    \caption{}
    \label{fig:subfig3}
  \end{subfigure}
  \hfill
  \begin{subfigure}[b]{0.49\textwidth}
    \includegraphics[width=\textwidth]{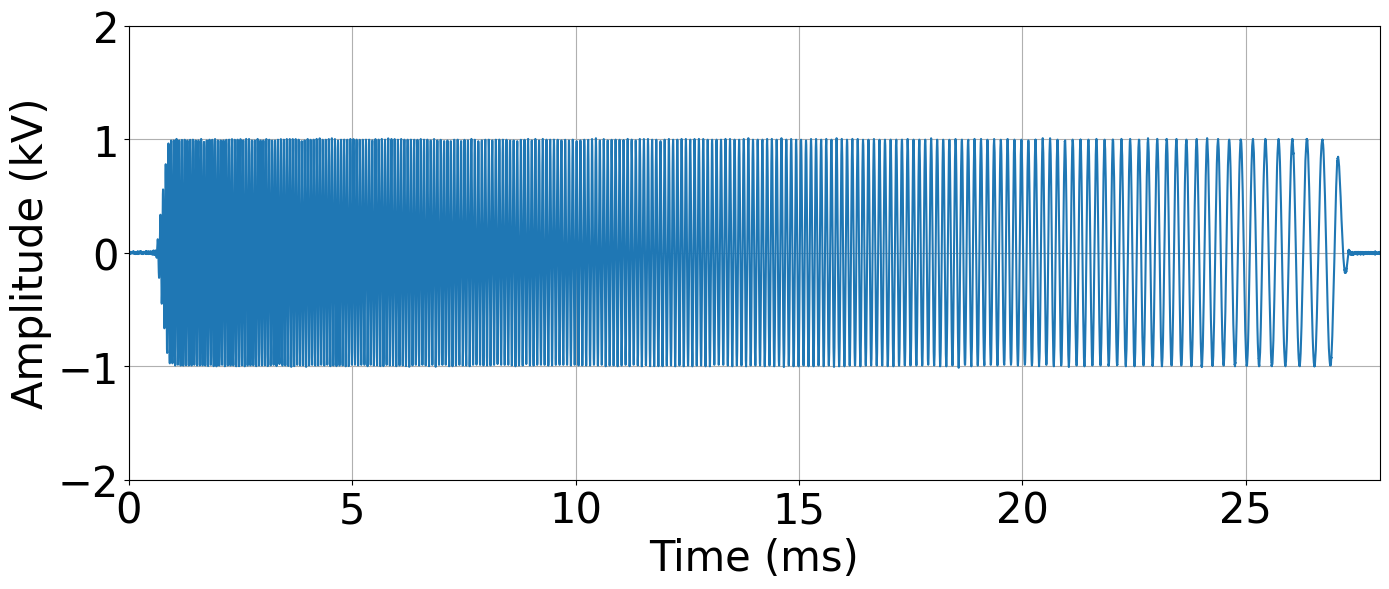}
    \caption{}
    \label{fig:subfig4}
  \end{subfigure}
  \caption{(a) Input waveform of channel A at the first iteration and (b) the corresponding output waveform. (c) Input waveform of channel A at the final iteration and (d) the corresponding output waveform. The target amplitude was set to 1 kV. A frequency sweep from 16.7 kHz to 2.5 kHz over 26.7 ms was used.}
  \label{fig:optimization_matrix}
\end{figure}

Figure \ref{fig:optimization_matrix} shows the results of the start and end of the optimization process for one of the channels. All eight transformers were connected to the 3-meter-long decelerator. After 53 iterations, the envelopes are flat within 2\% and the amplitudes are within 1\% of the target amplitude of 1 kV.

\vspace{1cm}

Note that in Figure \ref{fig:optimization_matrix} 'input waveform' refers to the waveform delivered by the Moku Pro to the audio amplifier. The results shown in Figure \ref{fig:optimization_matrix}b were obtained with all eight transformers connected to the decelerator and therefore include the effects of the capacitive coupling between the channels. Consequently, these results reflect the collective behavior of the system rather than the isolated response of a single transformer, and cannot be directly compared to the single transformer measurements shown in Figure \ref{fig:bodeplot}.

\newpage

\begin{figure}[H]
    \centering
    \includegraphics[width=1\linewidth]{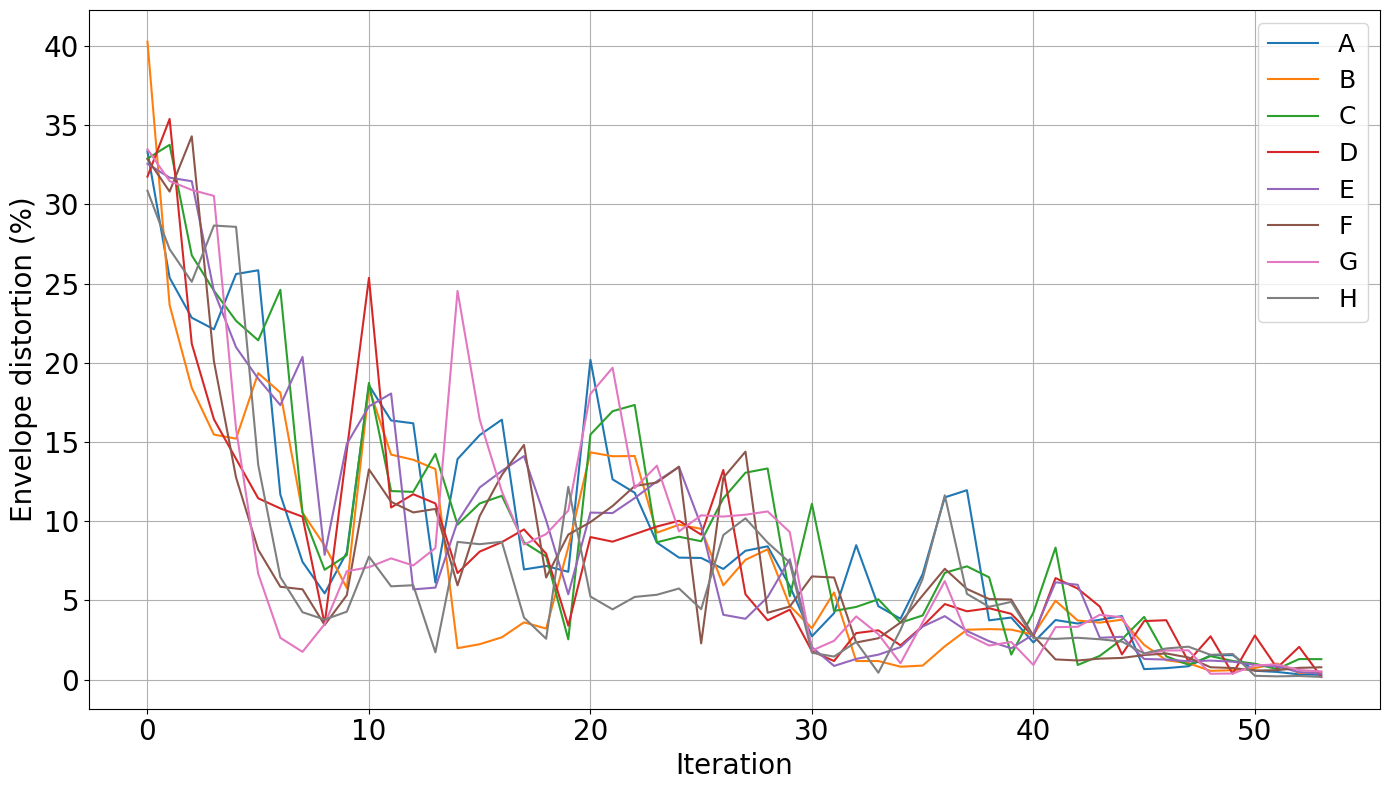}
    \caption{Envelope distortion of the transformer output waveforms as a function of the number of iterations for each channel (A-H).}
    \label{fig:envelope_flatness_full_dataset_1_v2}
    
\end{figure}

Figure \ref{fig:envelope_flatness_full_dataset_1_v2} shows the optimization process in more detail, plotting the envelope distortion as a function of the number of iterations for each channel. In this case the envelope tolerance starts at 20\% and is scaled down to 2\% over the course of 50 iterations. The effect of the gradual reduction of the acceptable tolerance can be seen by the several sharp increases in the envelope distortion throughout the first 50 iterations. When the envelope distortion is below the tolerance at a given iteration, the feedback system begins to optimize the amplitudes. As a result, the envelope distortion increases again due to the capacitive coupling between the channels, giving rise to the observed sharp peaks. 

\newpage

\begin{figure}[H]
    \centering
    \begin{minipage}[t]{0.49\textwidth}
        \centering
        \parbox[t]{\linewidth}{%
            \makebox[0pt][l]{(a)}\\[0.5em]
            \includegraphics[width=\linewidth]{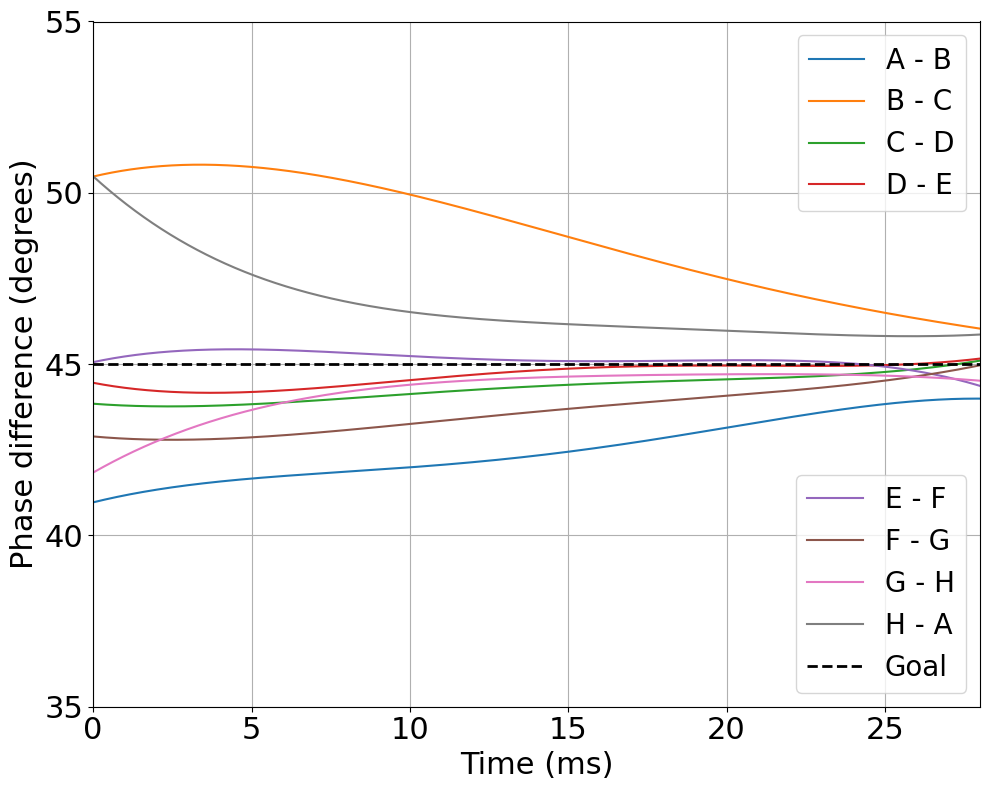}%
        }
    \end{minipage}
    \hfill
    \begin{minipage}[t]{0.49\textwidth}
        \centering
        \parbox[t]{\linewidth}{%
            \makebox[0pt][l]{(b)}\\[0.5em]
            \includegraphics[width=\linewidth]{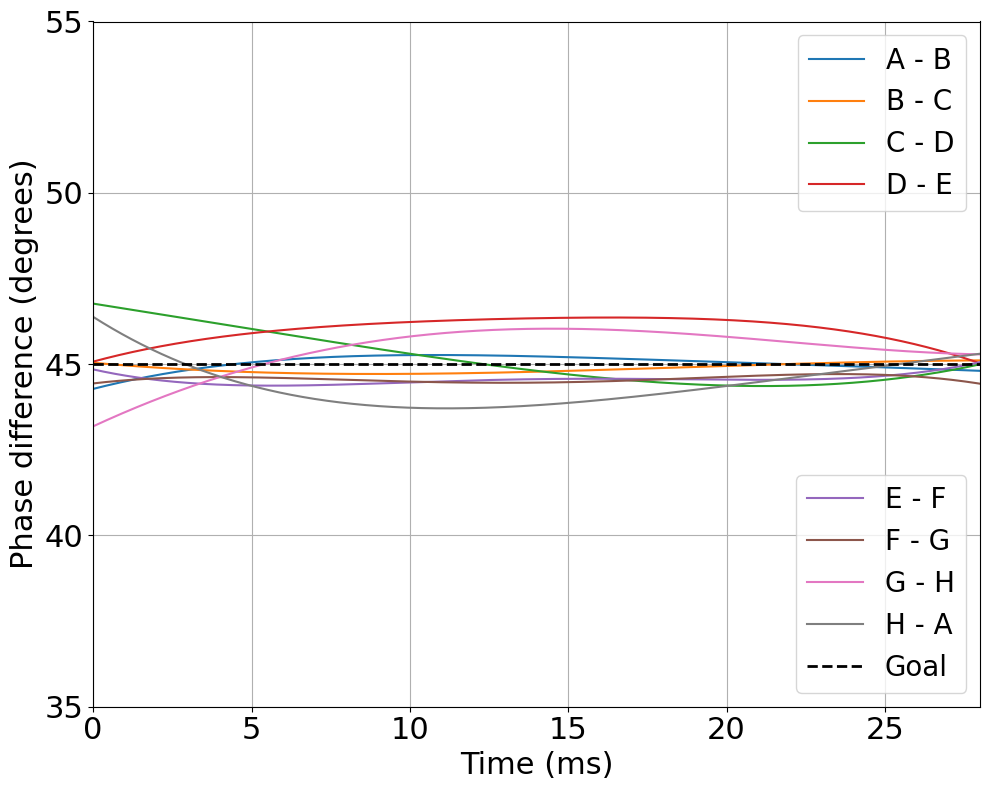}%
        }
    \end{minipage}

    \vspace{1em}

    \begin{minipage}[t]{0.7\textwidth}
        \raggedright
        \vspace{1em}
        \small
        \caption{(a) Phase differences of adjacent channels before the phase optimization. 
        (b) Phase differences of adjacent channels after optimization. (c) Final delays for each channel, starting from zero at the first iteration.} 
        \label{fig:delays_and_phases}
    \end{minipage}
    \hfill
    \begin{minipage}[t]{0.2\textwidth}
        \centering
        \vspace{1em}
        \parbox[t]{\linewidth}{%
            \raisebox{0pt}[0pt][0pt]{\makebox[0pt][l]{\hspace{-1em}(c)}}%
            \vspace{0.5em}
            
            \footnotesize
            \renewcommand{\arraystretch}{0.8}
            \begin{tabular}{|c|c|}
                \hline
                Channel & Delay ($\mu s$) \\
                \hline
                A & 0.98 \\
                \hline
                B & 1.41 \\
                \hline
                C & 0.50 \\
                \hline
                D & 0.14 \\
                \hline
                E & 0.55 \\
                \hline
                F & 0.81 \\
                \hline
                G & 0.82 \\
                \hline
                H & 1.47 \\
                \hline
            \end{tabular}
        }
    \end{minipage}
\end{figure}

Figure \ref{fig:delays_and_phases} shows the delays needed on each channel to optimize the phase differences between adjacent channels. The right panel shows the phase differences as a function of time after the optimization process for the 1 kV waveforms with a frequency sweep from 16.7 kHz down to 2.5 kHz.

Note that the phase differences in Figure~10(a) deviate significantly more from 45 degrees than Figure~\ref{fig:bodeplot} would suggest. These deviations are mostly caused by differences in delays of the audio amplifiers, not so much by differences in transformer performance.

\newpage

\begin{figure}[H]
    \centering
    \includegraphics[width=1\linewidth]{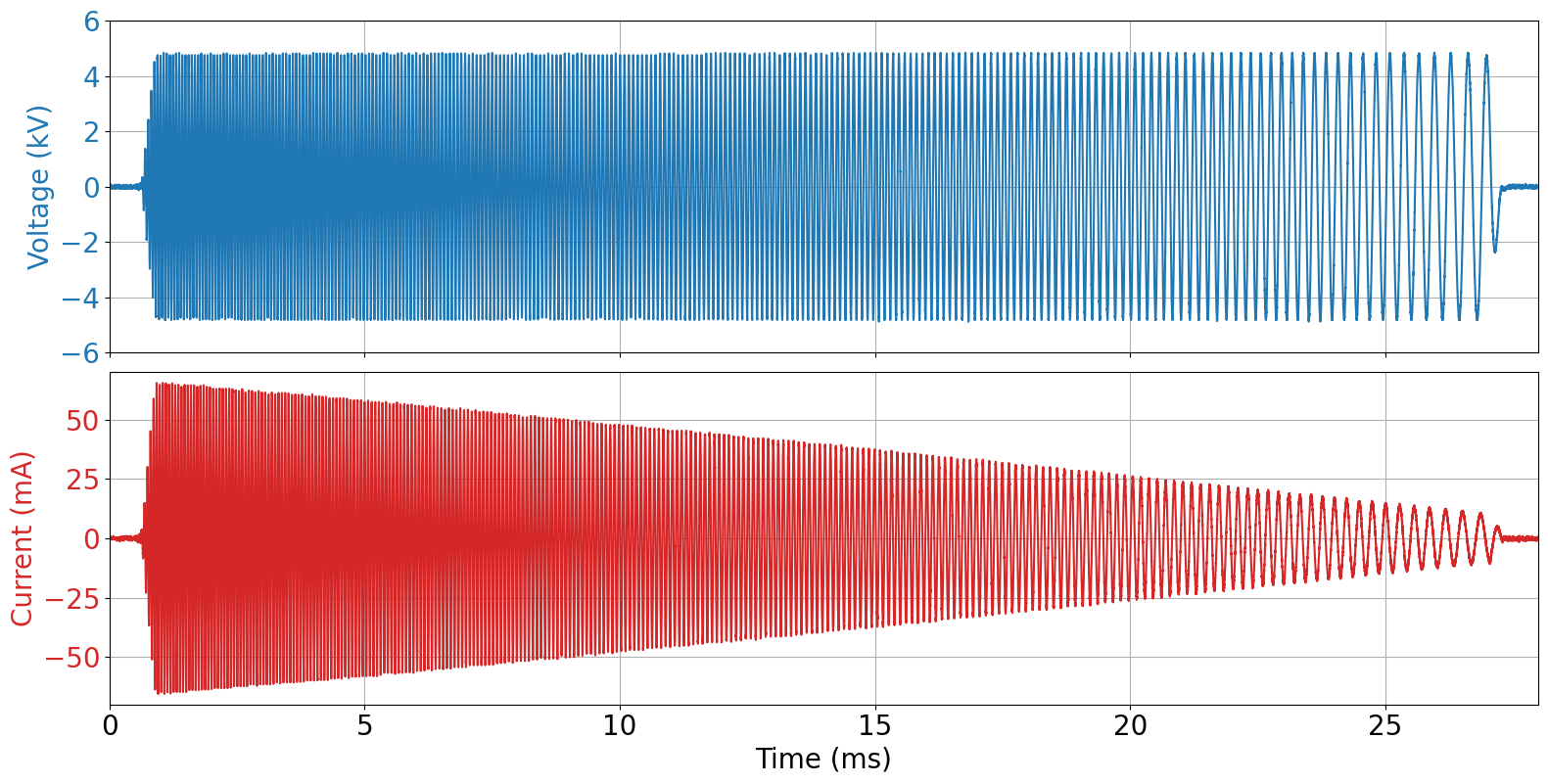}
    \caption{Output voltage and corresponding output current of one of the transformers connected to the 3 meter long decelerator after optimization with the target amplitude set to 5 kV and a frequency sweep from 16.7 kHz to 2.5 kHz.}
    \label{fig:voltage_current}
\end{figure}

After the optimization procedure at 1 kV, the amplitude is scaled up, after which only minor adjustments are needed. Figure~\ref{fig:voltage_current} shows the end result of this final optimization stage. The waveforms of all eight channels connected to the 3 meter long decelerator were optimized but only one channel is shown.

\subsection{Waveform fidelity}
\label{sec:waveform_fidelity}

\begin{figure}[H]
    \centering

    \begin{subfigure}{\textwidth}
        \begin{picture}(0,0)
            \put(-15,270){\textbf{(a)}}
        \end{picture}
        \includegraphics[width=\linewidth]{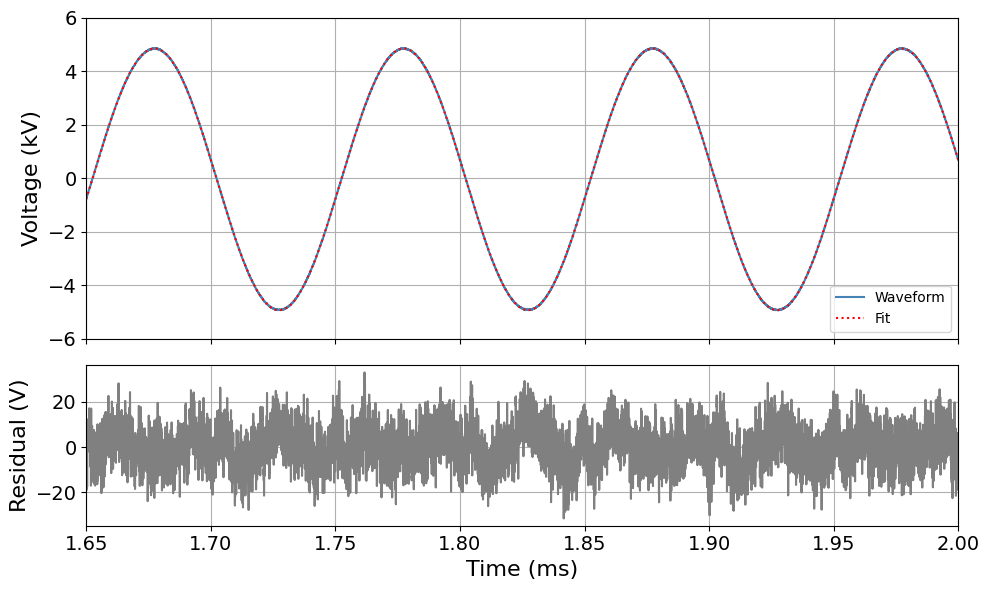}
    \end{subfigure}

    \vspace{2em} 

    \begin{minipage}[t]{0.6\textwidth}
        \begin{subfigure}{\linewidth}
            \begin{picture}(0,0)
                \put(-15,173){\textbf{(b)}}
            \end{picture}
            \includegraphics[width=0.99\linewidth]{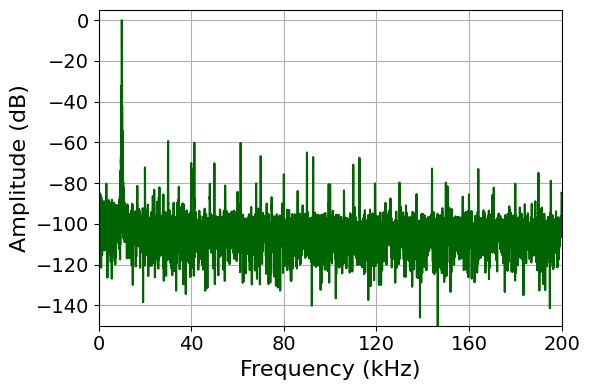}
        \end{subfigure}
    \end{minipage}
    \hfill
    \begin{minipage}[b]{0.35\textwidth}
        \raisebox{3.4cm}[0pt][0pt]{  
            \begin{minipage}[t]{\linewidth}
                \hspace{1cm}
                \begin{picture}(0,0)
                    \put(-20,75){\textbf{(c)}}
                \end{picture}
                \vspace{1em}
                \raggedright
                \fontsize{12}{9.6}\selectfont
                \begin{tabular}{|c|c|}
                    \hline
                    Channel & THD (\%) \\
                    \hline
                    A & 0.16 \\
                    \hline
                    B & 0.19 \\
                    \hline
                    C & 0.16 \\
                    \hline
                    D & 0.20 \\
                    \hline
                    E & 0.18 \\
                    \hline
                    F & 0.22 \\
                    \hline
                    G & 0.18 \\
                    \hline
                    H & 0.23 \\
                    \hline
                \end{tabular}
                
            \end{minipage}
        }
    \end{minipage}
        
    \caption{(a) Section of the output waveform of transformer A connected to the decelerator. The voltage amplitude is 5 kV at 10 kHz. A sinusoidal fit to the waveform is made and the difference between the fit and the waveform is plotted. (b) The corresponding power spectrum determined at a sampling frequency of 45.7 MHz. (c) The total harmonic distortion (THD) for the first 10 harmonics for each channel using the same waveform parameters.}
    \label{waveform_fidelity}
\end{figure}

One advantage of transformers as opposed to high-voltage amplifiers, is that they tend  to filter out frequency components significantly outside of their operational frequency range, since there is very little gain at those frequencies. The "purity" of the output waveforms, here referred to as waveform fidelity, can be characterized is several ways. One common metric used to quantify the waveform fidelity is the total harmonic distortion (THD) of the signal. Additionally, the measured signal can be compared to its theoretical ideal counterpart. Figure \ref{waveform_fidelity} shows the waveform fidelity for a voltage amplitude of 5 kV at 10 kHz.

\subsection{Operation at 10 kV amplitude}
\label{sec:10kv}

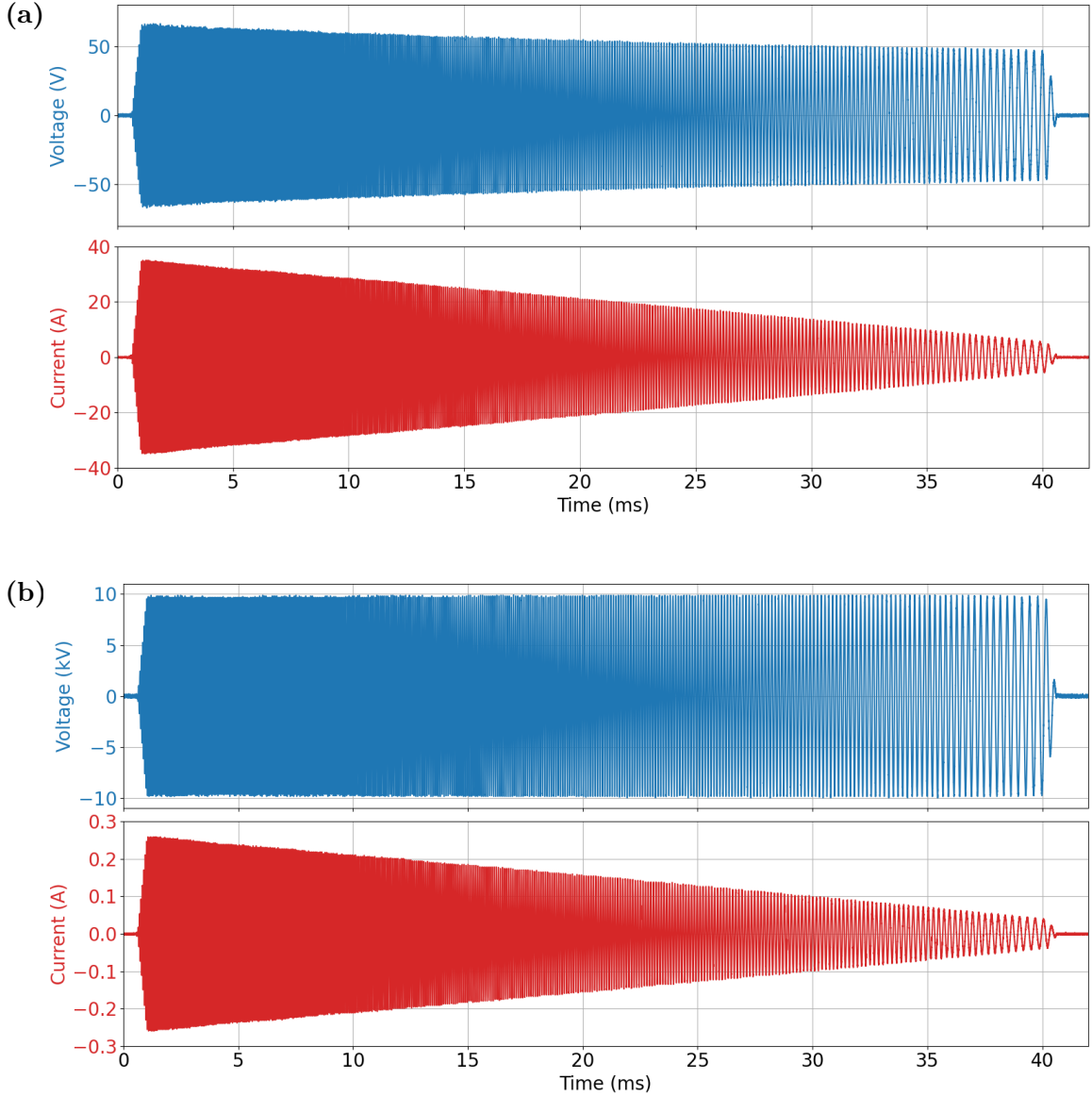
\begin{figure}[H]
    \centering

    \begin{subfigure}{\textwidth}
        \begin{picture}(0,0)
            \put(-12,200){\textbf{(a)}} 
        \end{picture}
        \includegraphics[width=0.9\linewidth]{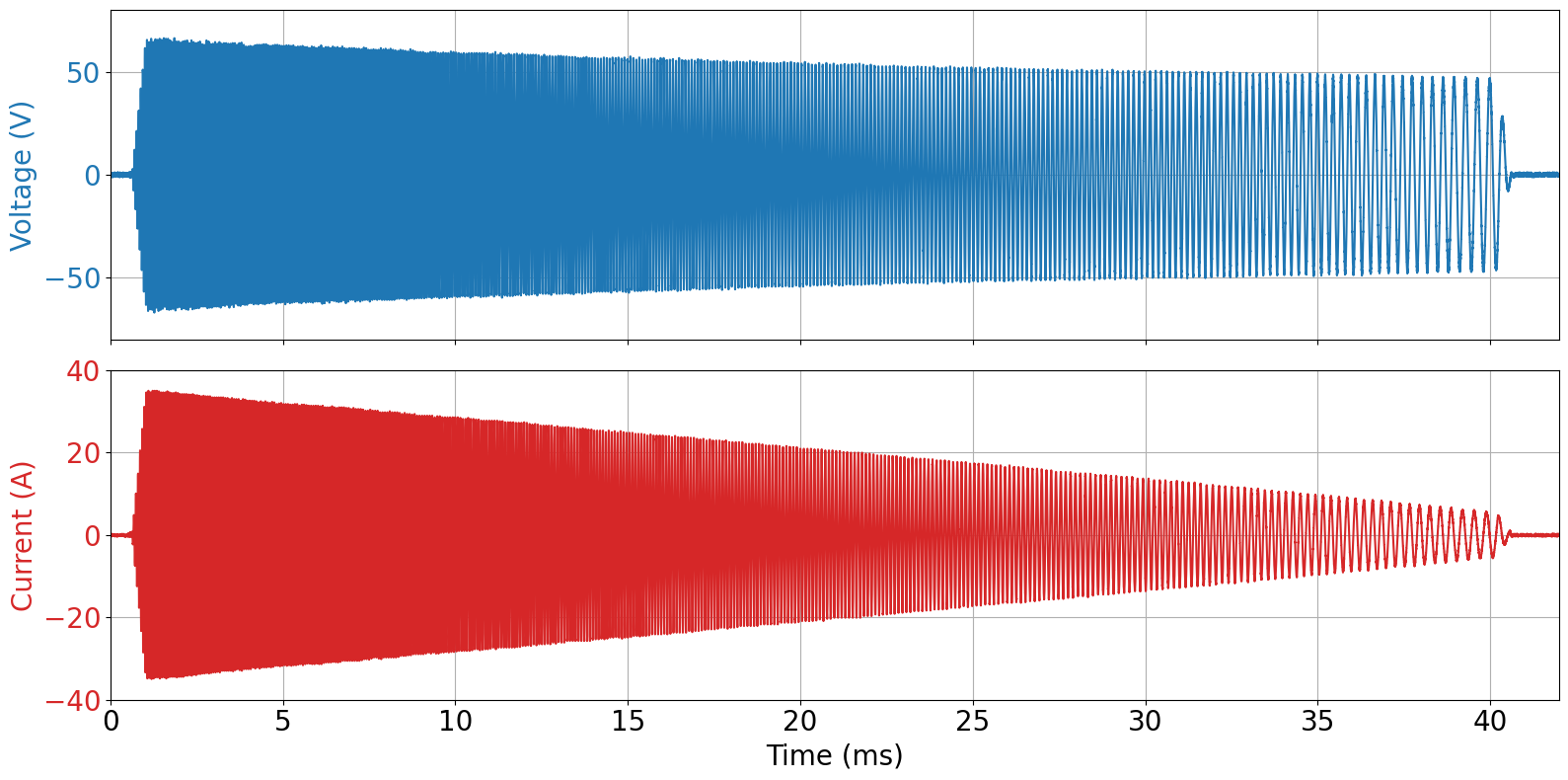}
    \end{subfigure}

    \vspace{2em}

    \begin{subfigure}{\textwidth}
        \begin{picture}(0,0)
            \put(-12,200){\textbf{(b)}} 
        \end{picture}
        \includegraphics[width=0.9\linewidth]{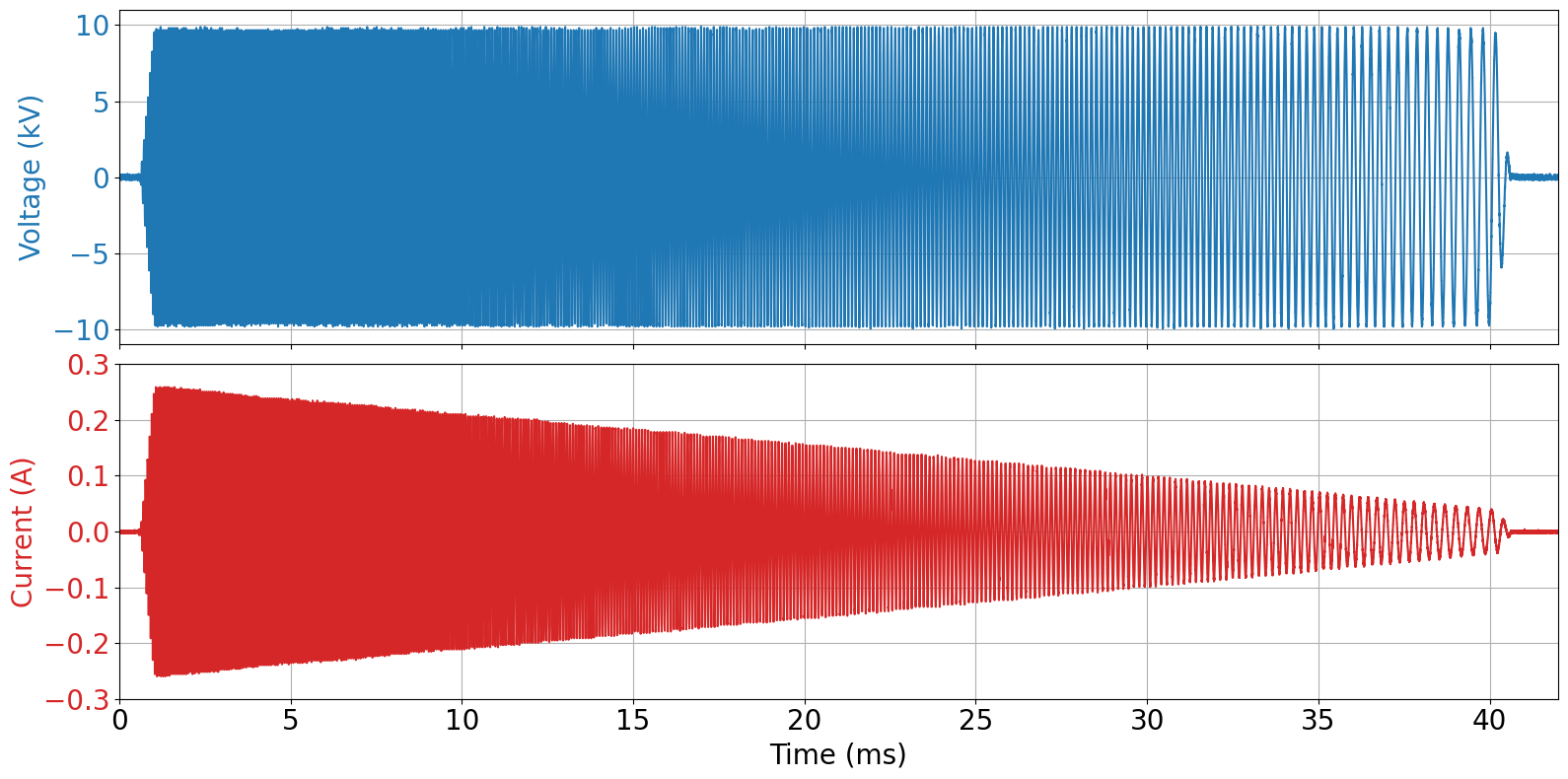}
    \end{subfigure}

    \caption{(a) Voltage across the transformer's input terminals and the corresponding current, and (b) the output voltage and corresponding current of a single encapsulated transformer driven by a frequency sweep from 16.7 kHz to 2.5 kHz in 40 ms on a 200 pF test load. The input waveform is optimized using the feedback system.}
    \label{10 kV}
\end{figure}

Figure \ref{10 kV} shows that a single encapsulated transformer can reach the desired 10 kV on a 200 pF test load using a 40 ms long frequency sweep from 16.7 kHz to 2.5 kHz at a repetition rate of 10 Hz. Here a 40 ms long waveform is chosen since this corresponds to the 4.5 meter long decelerator and it is a more stringent test of the hardware.

\section{Discussion}
\label{sec:discussion}

This section discusses the main results and limitations of the experimental setup.

The envelope distortion achieved over the target frequency range following feedback is below 2\%, and the amplitude deviates less than 1\% from the target value. These remaining small deviations reflect measurement noise, since also without feedback, the measured voltage amplitudes fluctuate by approximately 1\%.

The main limitation in the phase feedback originates from the finite time resolution of both waveform generation and measurement. The lookup table used for the waveform generation can contain at most 65536 data points, which for a 40 ms long waveform corresponds to a phase resolution of approximately 4 degrees at 16.7 kHz. This is insufficiently accurate to use in the phase feedback system. Similarly, the measurement system, operating at the maximum sample rate of 1.25 MSa/s for all four channels, exhibits comparable limitations. For this reason only the first millisecond of the waveforms are used to determine the phase differences, and the phase corrections are implemented via adjustments of the trigger levels rather than through modifications of the the lookup tables.

Since the waveforms are only shifted in time, differences in transfer functions between the channels with regard to the phase cannot be compensated for. However, the transfer functions of the transformers are nearly identical as shown in Figure~\ref{fig:bodeplot} and the largest phase difference resulting from differences in transfer functions is less than 0.5 degrees. The main contribution to the phase differences arise from differences in the delays of the amplifiers. These differences are several $\sim\mu s$, resulting in phase differences of $\sim 10$ degrees, which the feedback system compensates for. 

The achievable voltage amplitude is currently limited by the audio amplifiers. Although the amplifiers employed in the present setup are shown to be sufficient for reaching the 10 kV goal, they are operating near their limit. For this reason, using amplifier models that are capable of delivering a larger power output might be desirable for future research. However, the amplifier limit was reached when applying 10 kV across a 200 pF test load, whereas the 3 meter long decelerator has a capacitive load of about 107 pF, which significantly reduces the required input current.

\section{Conclusions and outlook}
\label{sec:conclusions_outlook}
Due to the demanding specifications of the high-voltage source required for the operation of a traveling-wave Stark decelerator (TWSD), in regards to voltage amplitude, current output, bandwidth, and waveform fidelity, no commercially available voltage source suffices.   
In this work the development and performance of a high-voltage source designed for the operation of a $\sim$ 4 m long TWSD to decelerate heavy polar molecules is presented. 
The described system is capable of producing voltage waveforms in the frequency range of 500 Hz to 16.7 kHz with an amplitude of up to 10 kV on a 200 pF load. 
The waveform fidelity was determined to be less than 0.25\% at 10 kHz, significantly better than that of the previously used TREK amplifiers and high-voltage sources used by other groups.
The feedback system can effectively compensate for any distortions of the waveforms, making the performance of the high-voltage source mostly independent of the details of the system. 
On the 3 meter long decelerator, the waveforms can be made to be flat within 2\%, are within 1\% of the set voltage amplitude and have a phase difference between adjacent channels within 2 degrees of the 45 degree goal.  The achievable accuracy is mostly limited by measurement noise caused by the limited voltage and time resolution. Ongoing simulation work aims to quantify the extent to which these remaining imperfections lead to molecular losses by comparing simulation results to measurements in future work. 
Components of the system that are most likely to require servicing are either made in-house or are relatively inexpensive and readily available. This reduces reliance on specialized external suppliers and enables repairs to be carried out quickly and at low cost when needed. 



\begin{acknowledgments}
\label{sec:acknowledgements}
We acknowledge the support of O.C. Dermois at the early stages of this project. We acknowledge funding from the Dutch Research Council (NWO) through grants XL21.074 and VI.C.212.016.
\end{acknowledgments}

\bibliography{HVgeneration.bib}
\end{document}